\documentclass[%
reprint,
 amsmath,amssymb,
 aps,
pra,
]{revtex4-1}

\usepackage{graphicx}% Include figure files
\usepackage{dcolumn}% Align table columns on decimal point
\usepackage{bm}% bold math
\begin{document}

%\preprint{APS/123-QED}

\title{Loading of a surface-electrode ion trap from a remote, precooled source}
\author{Jeremy M. Sage}
\email{jsage@ll.mit.edu}
\affiliation{Lincoln Laboratory, Massachusetts Institute of Technology, Lexington, MA, 02420}
\author{Andrew J. Kerman}
\affiliation{Lincoln Laboratory, Massachusetts Institute of Technology, Lexington, MA, 02420}
\author{John Chiaverini}
\affiliation{Lincoln Laboratory, Massachusetts Institute of Technology, Lexington, MA, 02420}
\email{john.chiaverini@ll.mit.edu}

\date{\today}

\begin{abstract}
We demonstrate loading of ions into a surface-electrode trap (SET) from a remote, laser-cooled source of neutral atoms.  We first cool and load $\sim$ $10^6$ neutral $^{88}$Sr atoms into a magneto-optical trap from an oven that has no line of sight with the SET.  The cold atoms are then pushed with a resonant laser into the trap region where they are subsequently photoionized and trapped in an SET operated at a cryogenic temperature of 4.6 K.   We present studies of the loading process and show that our technique achieves ion loading into a shallow (15 meV depth) trap at rates as high as 125 ions/s while drastically reducing the amount of metal deposition on the trap surface as compared with direct loading from a hot vapor.  Furthermore, we note that due to multiple stages of isotopic filtering in our loading process, this technique has the potential for enhanced isotopic selectivity over other loading methods.  Rapid loading from a clean, isotopically pure, and precooled source may enable scalable quantum information processing with trapped ions in large, low-depth surface trap arrays that are not amenable to loading from a hot atomic beam.

\end{abstract}

\pacs{Valid PACS appear here}

\maketitle

\section{Introduction}

Systems of trapped atomic ions are promising candidates for large-scale quantum information processing (QIP) \cite{Leibfreid:IonQIPReview:RevModPhys:03, Blatt:Wine:Nat08} due to the ions' demonstrated long coherence times, strong ion-ion interactions, and the existence of cycling transitions between internal states of ions for measurement.  Using trapped ions, single-qubit gates \cite{NIST:HifiMicrogate:12}, two-qubit gates \cite{Inns:HiFi2qubit:NatPhys:08}, and qubit state readout \cite{Oxford:HiFiRdout:PRL:08} have all been performed with fidelities exceeding those required for fault-tolerant QIP using high-threshold quantum error correction codes \cite{RaussHarr:HiThreshCodes:PRL:07}.  However, despite the promise shown by trapped ions, there are still many challenges that must be addressed in order to realize a practically useful quantum processor.  Chief among these is scaling up the number of simultaneously trapped ions while maintaining the ability to control and measure them individually with high fidelity.  Scaling up will require rapid, defect-free loading into large arrays.  Using the standard methods of ion loading by photoionization \cite{Drewson:PI:APB:00,Inns:PI:APB:01} or electron bombardment of neutral atoms from a hot vapor subjects the trap electrodes to surface contamination and, in the case of surface-electrode traps (SETs) \cite{NIST:SET:QIC:05, NIST:SET:PRL:06}, to possible electrical shorts due to metal deposition.  Additionally, these standard methods, including the use of backside loading \cite{NIST:Bksideloading:APL:09, Sandia:Bksideloading:arXiv:10, Bksideloading:GTRI:NJP:11}, will likely prove insufficient in their ability to perform isotopic filtering at the level required to have a manageably small number of defects (i.e., undesirable ion isotopes) in a large trap array.

Another major challenge is increasing the speed of all ion qubit operations without sacrificing fidelity.  Meeting this challenge will in general require shrinking the size of the individual ion traps.  However, this will require making the traps shallower to avoid electrical breakdown and trap instability \cite{NIST:SET:QIC:05}.  Loading directly from a hot source will prove inefficient when trap depths become comparable to or smaller than the mean energies of the hot atoms.  In addition, there have been many experiments that have observed heating of the external degrees of freedom of trapped ions at rates that scale as $d^{-4}$ \cite{Turchette:d4heating:PRA:00}, where $d$ is the distance from the ion to the electrodes and is proportional to the trap size.  This heating is larger than one would expect to arise from Johnson noise from the trap electrodes, and its origin is not definitively known.  It is problematic because significant heating of the ions' external degrees of freedom leads to infidelity in two-qubit gate operations \cite{NIST:ExpIssueswithIons:JresNIST:98}.  There is evidence that the trap electrode surfaces play a crucial role in the heating mechanism and that improving their quality and cleanliness leads to dramatic reductions in the anomalous heating rates observed \cite{NIST:ExpIssueswithIons:JresNIST:98, Chaung:CryoHeatRates:PRL08, NIST:Arclean:arXiV:12}.  Specifically, there is evidence that metal deposition onto the trap electrodes from the source used to load ions leads to an increase in the heating rate \cite{Turchette:d4heating:PRA:00, MIT:HeatingvsT:PRL:08}.  It has also been conjectured that the same surface contamination that causes heating leads directly to fluctuations of the motional frequencies of the ion trap, leading to further reduction of two-qubit gate fidelity \cite{NIST:Microwave2qubit:Nat:11}.  Thus, it is highly undesirable to coat the trap electrode surfaces with contaminants from the hot atomic beams used in conventional ion loading processes.

In this work we demonstrate a technique for loading ions that eliminates the contamination of electrode surfaces by hot atoms, is highly isotopically selective, and is highly efficient at loading shallow traps.  We begin by precooling neutral $^{88}$Sr atoms in a magneto-optical trap (MOT) from a hot oven that has no line-of-sight with our ion trap to prevent contamination of the electrodes.  We then push the cold atoms from the MOT toward the ion trapping region with a resonant laser, thus generating a pulsed cold atomic beam.  After a $\sim$ 1 ms flight time, the atoms reach the ion trap where they are photoionized.  Using this technique, we measure an ion loading rate into a low-depth SET which is comparable to or exceeds that of any rate previously reported using direct loading from a hot oven source, even when loading into deep, macroscopic (e.g., four-rod) ion traps.  Previous work demonstrating ion loading from pre-cooled and trapped neutral atoms in a MOT achieved higher rates  \cite{Chuang:MOTionload:PRA:07}, but in that case the MOT was collocated with the ion trap and thus contamination of the trap surfaces from the hot oven could not be avoided.

We note that our technique offers a potentially large improvement of isotopic selectivity over conventional ion loading methods.  The MOT is an intrinsically efficient isotopic filter, \cite{Lu:IsotopeAnalyses:Sci:99, ArgNatLab:CaMOTIsotopeSelect:PRL:04, Ko:SrMOTIsotpeSelect:JOSAB:06} and the resonant push beam offers additional selectivity as the linewidths of the cold atoms are smaller than the atomic isotope shifts.  Photoionization of the cold, isotopically pure atomic beam, while offering yet a third stage of filtering \cite{Drewson:PI:APB:00, Oxford:IsotopeSelectLoading:PRA:04}, is less sensitive to ionizing the incorrect isotope due to power broadening of the photoionization transitions or Doppler broadening arising from relative misalignment of the atomic and photoionization (PI) laser beams.  The experiments reported here are performed in a cryogenic environment with ion trap chip temperatures as low as 4.6 K while loading is being performed, thus demonstrating the compatibility of the loading technique with low-temperature systems.

\section{Apparatus}

\begin{figure}

\includegraphics[width=\columnwidth]{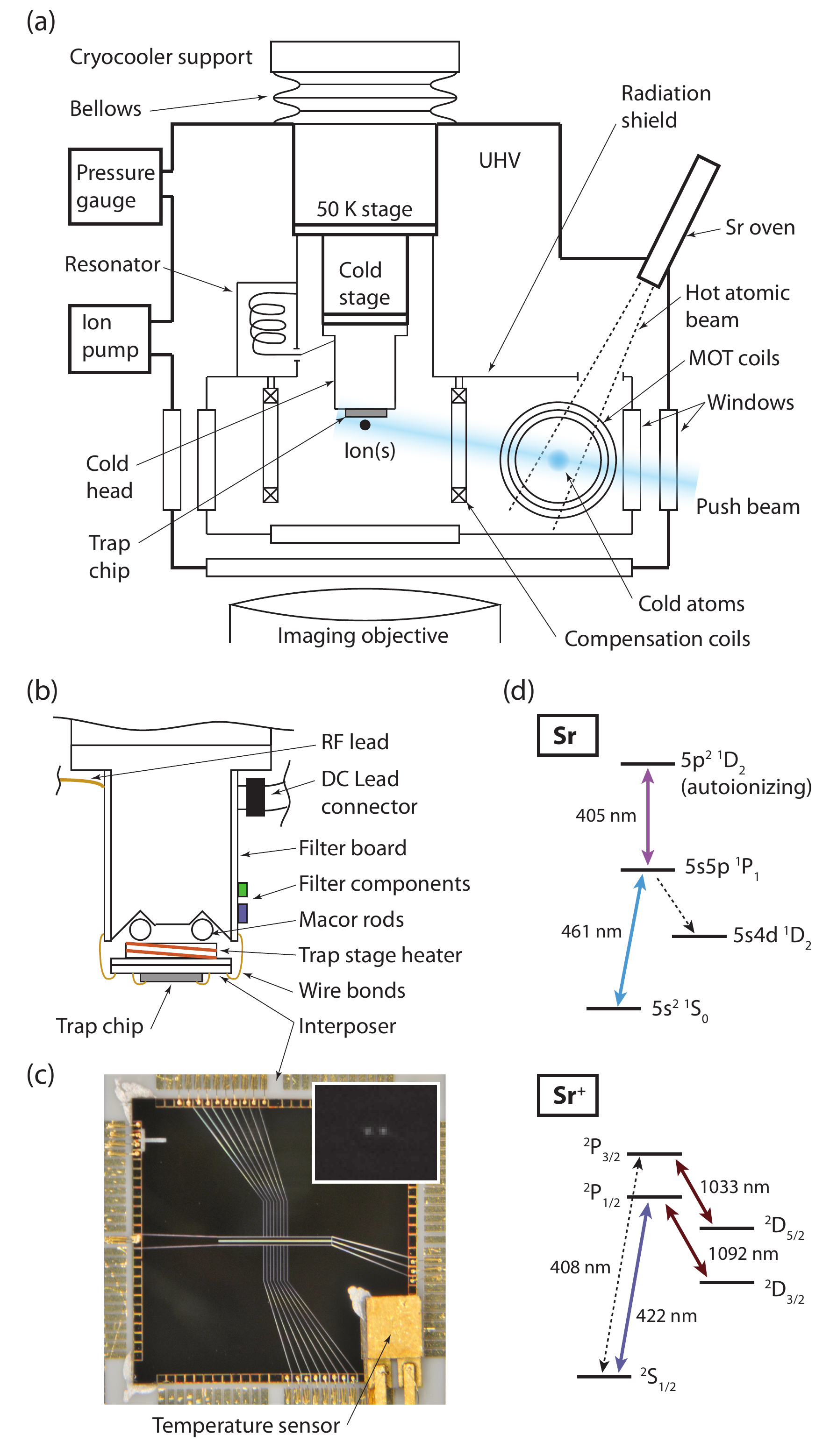}

\caption{\label{fig:apparatus} (Color online) Cryogenic apparatus and atomic systems.  (a) Diagram depicting the cryogenic UHV magneto-optical trap (MOT) and ion trap system (not to scale).  (b) Detail of cold head and trap stage (not to scale).  Control signals are routed from the lead connector, along the filter boards (through single-stage low-pass filters), onto the interposer, and onto the trap chip, with gold wire bonds used at each junction.  The radio-frequency (RF) signal is routed similarly, excepting the filter components.  Macor rods thermally isolate the trap stage from the cold head, allowing the former to be heated separately from the cold stage of the cryocooler.  (c) Photograph of the surface-electrode ion-trap chip (the chip is 1 cm on a side).  The trap is niobium on a sapphire substrate, and the patterned metal electrodes form a segmented linear Paul trap 50 $\mu$m from the chip surface.  The two RF electrodes are fed from the central lead on the left, and all other electrodes are DC control electrodes.  The inset shows a fluorescence image of two ions in this trap (the inter-ion distance is 6 $\mu$m).  (d) Relevant electronic level structures of neutral and singly ionized strontium.  Solid arrows represent transitions driven directly in this work and dotted arrows represent other relevant excitation or decay paths.}

\end{figure}
\subsection{Cryostat and Vacuum System}
A schematic of the apparatus is shown in Fig.~\ref{fig:apparatus}(a).  The cryostat consists of a two-stage closed-cycle Gifford-McMahon cryocooler  and a sample mount that is mechanically connected to the cryocooler only via a flexible rubber bellows (Janis Research Company SHI-4XG-15-UHV).  The volume between the sample mount and the cold head of the cryocooler is filled with helium exchange gas to provide a thermal link to cool the sample mount while isolating it from the cold head's vibrations.  The cryocooler is supported by a frame mounted to the ceiling and the sample mount is attached via conflat seal to an ultra high vacuum (UHV) chamber mounted on an optical table.  As a result, vibrations from the cryocooler are well isolated from the UHV system and from the table.    In this configuration, the first stage of the sample mount has a specified base temperature of 50~K with a 45 W cooling power and the second (``cold") stage has a specified base temperature of 3 K with a 1 W cooling power.  A radiation shield is attached to the first stage to prevent thermal radiation from the room temperature chamber walls and windows from heating the cold stage.  The shield has windows that provide optical access to the ion trap region and has holes to allow access for electrical wires and for vacuum pumping.  Optical access to and from the UHV chamber for laser beams and fluorescence imaging is provided through indium-sealed viewports.

High vacuum is achieved via a combination of a turbomolecular and a 55 $\ell$/s ion pump; to reach UHV, we rely mainly on cryopumping.  We do not bake any part of the apparatus and we achieve a pressure below $1 \times$ $10^{-9}$ Torr a few hours after the cryostat is turned on, as measured by an inverted magnetron gauge located near the first stage of the cryostat.  After about 1 week of cryopumping (with the ion pump on) we achieve a pressure below $1 \times$ $10^{-10}$ Torr; the pressure near the ion trap located at the colder second stage of the cryostat should be significantly lower than this value to do more effective cryopumping.

\subsection{Lasers}
Several sources of laser light are required to address the relevant transitions in strontium as shown in Fig.~\ref{fig:apparatus}(d).  Light is needed at 461 nm for the MOT, the neutral atom push beam, and the first step of photoionization; 422 nm for Doppler cooling of Sr$^+$; 405 nm for the second step of photoionization; 1092 nm for repumping ions decaying from the $P_{1/2}$ to the $D_{3/2}$ level; and 1033 nm to repump ions excited to the $P_{3/2}$ level by amplified spontaneous emission from the 405 nm laser that subsequently decay to the $D_{5/2}$ level.  The 461 nm light is generated by frequency-doubling a high power ($>$ 1 W) 922 nm tapered amplifier source which produces as much as 400 mW of 461 nm light at the output of the doubling cavity.  As much as 10 mW of 422 nm light is generated by frequency doubling a lower power 844 nm source (no tapered amplifier).  All other lasers are of the external-cavity grating-stabilized design except the 1092 nm laser which is a distributed feedback laser.

For laser frequency stabilization, we lock an additional laser at 852 nm to a cesium saturated absorption signal, and this laser is  used to stabilize the length of a Fabry-Perot ``transfer" cavity. The frequencies of the 461 and 422 nm lasers are then actively stabilized to a bandwidth $<$ 100 kHz by locking the 922 and 844 nm light to the transfer cavity.  The 1092 nm laser is stabilized by active feedback from a wavemeter measurement with a resolution of $\sim$~100 MHz providing stability on the order of this resolution.  We do not lock the 405 and 1033 nm lasers, relying on their passive stability which is sufficient for the experiments done here.  All the laser systems except the 405 nm laser are physically located in a separate room from the ion trap apparatus.  When located together, we found that acoustic noise from the cryocooler affected the lasers; in particular amplitude noise on the 461 and 422 nm systems arose due to disruption of the doubling cavity locks.  The laser light is transferred between rooms via polarization-maintaining single mode fiber.  All laser beams that are sent to the neutral atoms and ions can be switched on and off with either acousto-optic modulators or mechanical shutters, depending on the speed required.

\subsection{Cold Head and Ion Trap}
The cold head is a copper block that is mounted to the cold stage of the cryocooler and is detailed in Fig.~\ref{fig:apparatus}(b).  We indium solder a total of four 1 mm thick alumina filter boards to two sides of the cold head.  Each filter board is attached at its base to a 24-pin connector that mates with a connector bringing phosphor-bronze DC wires from room temperature.  The filter boards have patterned gold traces running along the board length to which we attach DC filter components with conductive epoxy to low-pass filter each DC line at a cutoff frequency of $\sim$ 10 kHz.  A copper wire lead coming from a radio-frequency (RF) resonator is soldered directly to an additional unfiltered trace on one of the filter boards to carry the RF signal to the ion trap chip.  An important function of this cold-head and filter-board configuration is to effectively heat sink all the DC and RF leads that go to the ion-trap chip.

The trap mount is a smaller copper piece that is mechanically attached to the cold head but is thermally isolated from it through the use of macor rods.  A nichrome heater wire is wrapped around the base of the trap mount to allow the temperature to be varied, and a 1-mm-thick alumina ``interposer" piece with patterned gold traces is indium-soldered to its bottom.  Cooling of the trap mount is achieved primarily through the $\sim$ 100 Au wire bonds that connect the interposer traces to the traces on the filter boards.  We use silver paint to fix the ion-trap chip to the alumina interposer (so that it can be easily removed when changing traps) and wire bonds from the interposer traces make all electrical connections to the trap electrodes.  A silicon-diode temperature sensor is attached with silver paint directly to the ion-trap chip surface.  The chip temperature gets as low as 4.3 K with all wires connected and with the lasers and the RF voltage turned off.  Applying the RF power at the level required for ion trapping ($\sim5$~mW) causes insignificant heating; however, turning on the lasers required for ion loading and trapping causes the temperature to increase to 4.6 K.  We have verified that we can raise the temperature of the trap chip with our trap mount heater to at least 100 K while keeping the cold head at 3.5 K, essential for maintaining a UHV environment that relies on cryopumping.

The ion Paul trap is an SET design with a linear geometry.  It is composed of 2 $\mu$m thick niobium sputtered onto a 430 $\mu$m-thick sapphire substrate.  Patterning features as small as 4 $\mu$m is done by optical lithography.  The chip, shown in Fig.~\ref{fig:apparatus}(c), is designed to trap ions 50 $\mu$m from the surface utilizing two RF electrodes and 18 DC electrodes to generate a pseudopotential for confinement along the radial directions and to allow for confinement along the axial trap direction in up to four separate, simultaneous trapping zones (seven total trapping zones).  Two additional DC electrodes running down the center of the trap are used for tilting the radial trap axes out of the plane of the chip so that all vibrational modes may be Doppler cooled with a single laser beam oriented parallel to the chip surface \cite{Oxford:TrapDesign:NJPhys:10}.  All DC electrodes are additionally used for micromotion compensation \cite{berkeland:micromotion:JAP:98}.

We operate the trap using only one trapping zone with an RF voltage amplitude of 35 V and all DC electrode voltages $<$ 10~V.  To generate the RF voltage, we use a helical resonator mounted inside the UHV chamber and attached to the radiation shield to step up a lower voltage signal derived from room temperature electronics outside the vacuum system.  The resonator has a resonance frequency of 27.5~MHz and a loaded Q value of 280 giving a voltage step-up of 50.  The radial trap frequencies are measured to be 3.0 and 3.7~MHz and the axial trap frequency is measured to be 640~kHz; the trap depth is 15~meV, determined by boundary element simulation and the measured frequencies.

\subsection{Imaging and Detection Systems}
Fluorescence from the the neutral MOT atoms is imaged onto a CMOS camera.  Integration of this signal is used to determine the MOT atom number.  Fluorescence from the ions is collected with a numerical aperture of 0.4 (\textit{f}/1.2) using a custom compound objective placed just outside the bottom window of the chamber.  An aperture is placed in the primary focal plane, and the light is refocused through a relay lens onto either a photomultiplier tube (PMT) or an electron-multiplying CCD camera.  The total magnification is 46x with an observed resolution of $\sim$ 1 $\mu$m.

\section{Cold Neutral Atom Source}
We cool and trap neutral $^{88}$Sr atoms in a MOT located remotely ($\sim$ 65 mm) from the ion trap region to avoid deposition of hot atomic vapor on the ion trap electrodes.  We load the MOT directly from an effusion oven that has no line of sight with the ion-trap chip to accomplish this.  Up to 100 mW of total laser power at 461 nm is available for the MOT and it is split into three 25-mm-diameter beams.  We do not use any repumping lasers because the $^1S_0\rightarrow$$^1P_1$ transition is nearly cycling; the $^1P_1$ state decays to the dark $^1D_2$ state at a rate $\gamma_{dark}=1.3\times$ 10$^3$~s$^{-1}$ which is much less than the rate for decay to the $^1S_0$ state, $\gamma_{P-S}=2\times 10^8$ s$^{-1}$ \cite{Ye:SrMOT:JOSAB:03}.  We do not employ a Zeeman slower for MOT loading but we note that doing so would yield higher atom number at the expense of greater experimental complexity.  The MOT magnetic field coils are mounted to the first stage of the cryostat and produce a gradient of 65 G/cm (in the axial direction) with 2 A of current; we observe no heating of the cold stage or trap chip when current is applied.  With our chosen field gradient and laser beam size, we find that a laser detuning of $-64$ MHz ($=-2\gamma_{P-S}/2\pi$) relative to the $^1S_0\rightarrow$$^1P_1$ resonance maximizes the MOT atom number.

\begin{figure}

\includegraphics[width=\columnwidth]{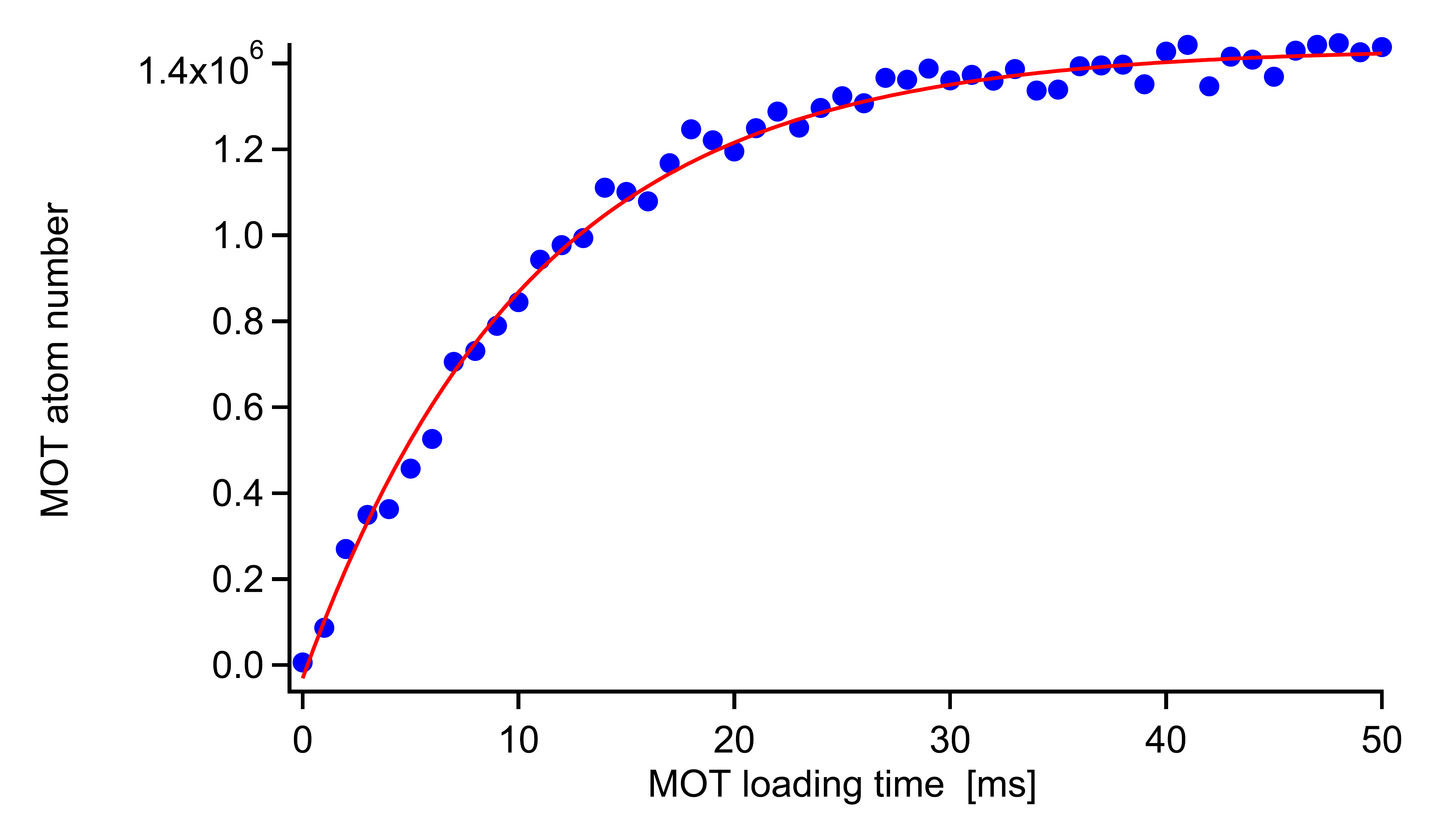}

\caption{\label{fig:MOTload} (Color online) Number of atoms trapped in the MOT as a function of MOT loading time.  These data were taken at an oven temperature of $425\,^{\circ}\mathrm{C}$ and a total MOT laser beam intensity of 40 mW/cm$^2$.  The line is a fit to a simple exponential (see text) with a characteristic time of $\tau=10.5$~ms and a steady-state atom number of $N_{ss}=1.4 \times 10^6$.}

\end{figure}

The atom number $N$ and the characteristic 1/\textit{e} MOT loading time $\tau$ are important factors in the attained ion loading rate.  In Fig.~\ref{fig:MOTload} we show a typical MOT loading curve; the data show the well-known MOT behavior where $N$ increases with load time until it saturates.  The value of $\tau$ is given by $\tau^{-1}=\alpha\gamma_{dark}+\gamma_{bkd}$, where $\gamma_{bkd}$ is the loss rate from the MOT due to collisions with background atoms and $\alpha$ is the fraction of the atomic population in the $^1P_1$ level, which is a function of the total laser light intensity.  In our MOT loading scheme, the hot atomic beam passes directly through the magneto-optical trapping region so that there is a high local density of energetic atoms which leads to a substantial background loss rate.   As a result, we find that $\alpha\gamma_{dark}$ and $\gamma_{bkd}$ are comparable in magnitude for the $375-425\,^{\circ}\mathrm{C}$ range of oven temperature and 20-40 mW/cm$^2$ range of total laser intensity we use here.  The steady-state atom number $N_{ss}$ (measured after a time $\gg \tau$) is given by $R_{\ell}\tau$, where $R_{\ell}$ is the MOT loading rate. The value of $R_{\ell}$ is a function of the oven temperature (atom flux) and the intensity of the trapping light, increasing in value as both are increased.  Fitting the data in Fig.~\ref{fig:MOTload} to the simple exponential loading function $N(t)=N_{ss}(1-e^{t/\tau})$, we find $N_{ss}=1.4\times10^6$ and $\tau=10.5$ ms.  The temperature of the cold atoms is determined to be 8 mK by measuring the expansion of the initially $\sim$ 2 mm diameter atom cloud (via imaging on the CMOS camera) with the MOT trapping beams blocked for a fixed time.

After loading the MOT for a time $t_{MOT}$, we switch off the MOT laser beams and immediately turn on the push laser beam, accelerating the entire cold atom cloud.  The push beam is resonant with the Sr $^1S_0\rightarrow$$^1P_1$ transition and is directed through the MOT along a line intersecting the center of the ion trap.  The push beam diameter is $\sim$~2~mm and its intensity is chosen by increasing the beam power until we observe no further improvement in the ion loading rate.  The push beam hits the surface of the ion trap chip; however we observe no heating of the chip or increase in excess ion micromotion (which would indicate charging of the dielectric) for the $\sim$ 1 mW of push beam power used.  We note that in principle one could avoid striking the chip with the pushing light by using two push beams that are not directly aimed at the ion trap but instead have a resultant \textit{k}-vector that points along this direction.  This would be useful if employing this technique for pushing atoms requiring shorter wavelength light (e.g., Ca, Yb) that may cause significant charging \cite{Oxford:CaCharging:APB:11, ChuangMIT:TrapCharging:JAP:11}.

The pushed cold atoms arrive in the ion trap region in a pulse delayed in time $\sim$ 1 ms after the push beam is turned on.  We simulate the trajectory of the atoms in the presence of the push light and the MOT magnetic field, taking into account the Zeeman and Doppler shifts (the effect of gravitational acceleration is negligible during the atom flight time).  We calculate a maximum atom speed of 72 m/s at the measured pulse arrival time, which corresponds to 2.3~meV of kinetic energy,  much less than the 15 meV trap depth.  From the simulation we also calculate the cold atom beam's transverse spatial divergence due to the initial velocity spread from the 8 mK MOT atoms and the recoil momentum kicks from the push light scattering; we find that the beam size increases from the 2 mm MOT diameter to a 4 mm diameter at the ion trap location and that the contribution to the atoms' kinetic energy from the transverse velocity is negligible.  Finally, we calculate that $<$ $5\%$ of the atoms are ``lost" from the cold beam due to pumping to the dark state by the push light during the 1 ms flight time.

\section{Ion Loading}
\subsection{Photionization and Doppler Cooling}
All laser beams used to address ions are focused to a diameter of $\sim$ 50 $\mu$m and travel parallel to the surface of the trap chip.  To photoionize the neutral atoms, a 461~nm laser beam resonant with the Sr $^1S_0\rightarrow$$^1P_1$ transition is spatially combined with a 405 nm laser beam in order to drive the two-step transition to the auto-ionizing $5p^2$ $^1D_2$ level \cite{Chiaverini:SrPI:arxiv:06, Mende:SrPI:JPhysB:95, Gill:SrPI:APB:07}.  The combined photoionization beams propagate roughly orthogonally to the cold neutral atom beam and parallel to the axial direction of the linear ion trap (``trap axis").  The 422 nm Doppler cooling beam, with a power of 200 $\mu$W, intersects the PI beams at the center of the ion trap at a $\sim\nobreak36^{\circ}$ angle to the trap axis.  The 1092 and 1033 nm laser beams used for repumping, with powers of 900 $\mu$W and 1.6 mW, respectively, are combined and overlap the other beams at the center of the ion trap.  We apply a 6 G magnetic field with compensation coils (mounted to the first stage of the cryostat) along the direction of the trap axis and at an angle to the 1092 nm linear laser polarization in order to prevent optical pumping into dark states \cite{Dehmelt:darkstates:JOSAB:85}.  We measure the presence of ions in the trap by monitoring their fluorescence from the Doppler cooling laser whose frequency is detuned by $\sim$ 150 MHz, an amount such that the measured fluorescence of the ion is approximately one half of the maximum value observed on resonance.  We experimentally find that the ion loading rate is only very weakly dependent on this detuning over a $\pm$ 10 MHz range around our chosen value.  With the Doppler cooling and 1092 nm repump laser on, we measure an average ion lifetime in our trap of $\sim$ 5 minutes.

\subsection{Measurement of Ion Loading Probability and Rate}
Measurements of ion loading rates are typically performed in relatively deep traps capable of stably trapping many ions \cite{Oxford:IsotopeSelectLoading:PRA:04, Drewsen:Ablationloadingrate:APB:07, Chuang:MOTionload:PRA:07, Schaetz:BaLoading:arxiv:11}.  This allows one to measure the rate of increase of the total fluorescence from the trap as multiple ions are loaded in order to extract the loading rate.  In our shallow trap, operating at a stability parameter of 0.4, we never observe more than two co-trapped ions, making this method insensitive to measurement of high loading rates.  As a result, we instead measure the probability to load a single ion $P_{ion}$ and from this calculate the loading rate $R_{ion}=P_{ion}/t_{\ell}$ where $t_{\ell}$ is the total time taken to load the ion trap with probability $P_{ion}$.  We obtain the value of $P_{ion}$ by attempting to load an ion from repeated single pulses of the cold atom beam and measuring the fraction of total pulses in which loading is successful.  The RF voltage to the trap is switched off for 100 $\mu$s between each pulse to eject ions from the trap with certainty in order to make successive loading attempts independent. Successful loading is determined by counting photons from the ion fluorescence with the PMT and comparing the result to a threshold value chosen to lie above what we measure when no ion is present in the trap.

\begin{figure}

\includegraphics[width=\columnwidth]{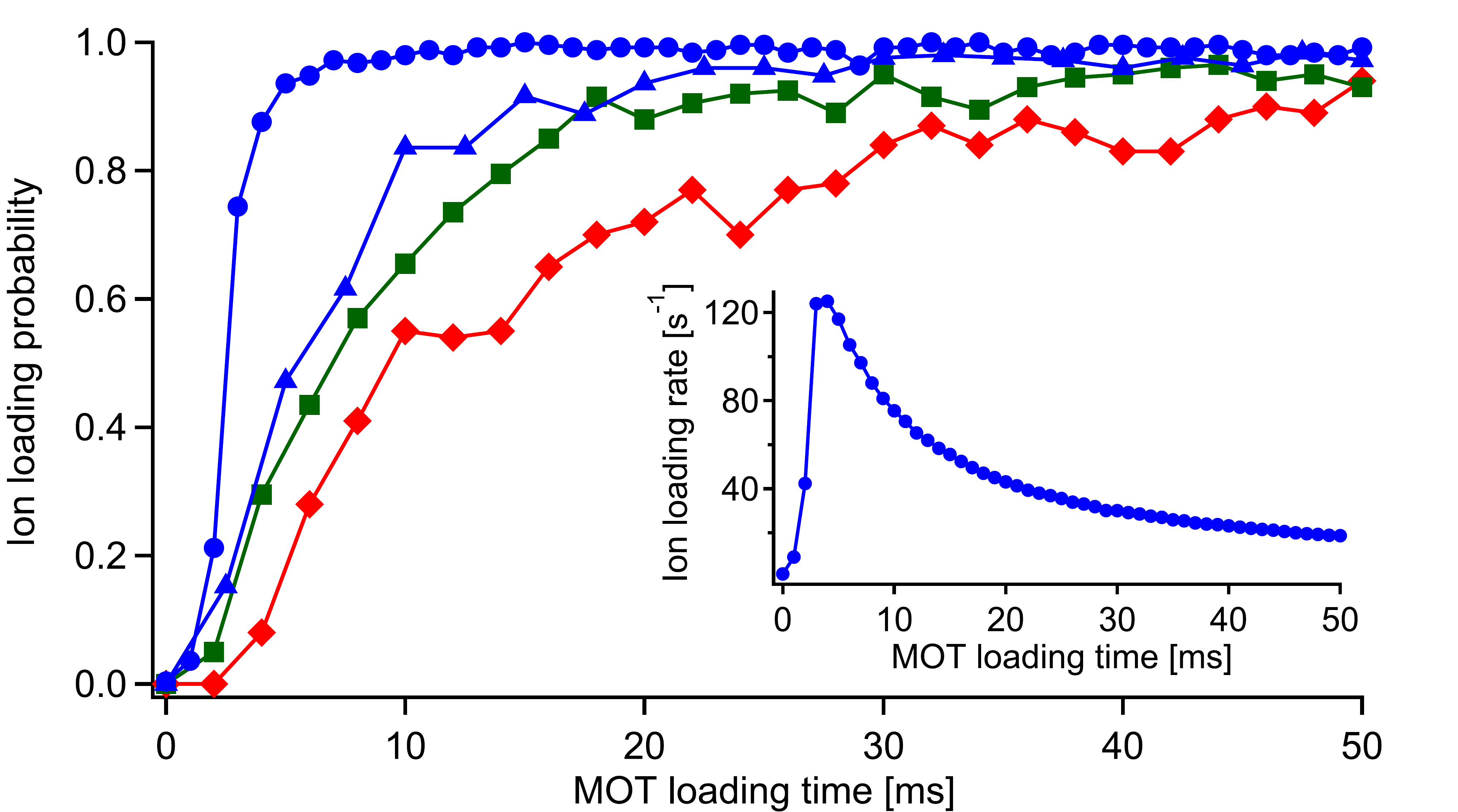}

\caption{\label{fig:ionload} (Color online) Ion loading probability and rate as a function of MOT loading time.  The upper set of points, the (blue) circles, are for an oven temperature $T_\mathrm{oven}=425^{\circ}$C and a total MOT beam intensity of 40 mW/cm$^2$.  The lower three sets of points are for varying $T_\mathrm{oven}$ at a lower MOT beam intensity of 23 mW/cm$^2$:  the (blue) triangles are for $T_\mathrm{oven}=425^{\circ}$C, the (green) squares are for $T_\mathrm{oven}=400^{\circ}$C, and the (red) diamonds are for $T_\mathrm{oven}=375^{\circ}$C.  The inset shows the ion loading rate as a function of MOT loading time obtained from the upper set of points in the main figure.  The data are taken with the push and photoionization (PI) lasers on for 3~ms after the MOT beams are turned off.  All lines are guides for the eye.}

\end{figure}

In Fig.~\ref{fig:ionload} we show $P_{ion}$ versus $t_{MOT}$ for three different oven temperatures and two different MOT laser powers.  The data are taken with 5 and 9 mW of power in the 461 and 405 nm PI beams, respectively.  The push and PI beams are switched on for a time $t_{ion}=3$~ms immediately after the MOT beams are switched off and the PMT measurement is subsequently gated for 10~ms once the push and PI beams are switched off.  The Doppler-cooling and repump beams are on continuously.  Each curve in Fig.~\ref{fig:ionload} shows that $P_{ion}$ increases with $t_{MOT}$ before saturating.  This is to be expected since the loading probability should increase with a larger number of cold atoms in each beam pulse.  Comparing the different curves in Fig.~\ref{fig:ionload}, we see that $P_{ion}$ increases with higher oven temperature and higher MOT laser power which is again expected because both lead to a measured increase in MOT atom number \cite{MOTDensitynote}.  As can be seen in the curve for $T_{\mathrm{oven}}$ = 425$^{\circ}$C and 40 mW/cm$^2$ total MOT beam intensity, we attain near unit probability for loading a single ion in $\sim$ 10 ms.  In the inset of Fig.~\ref{fig:ionload}, we plot $R_{ion}$ versus $t_{MOT}$ for this same oven temperature and MOT beam intensity.  The values of $R_{ion}$ are obtained using $t_{\ell}= t_{MOT}+ t_{ion}$.  We observe that the maximum rate is $R_{ion}$ = 125 ions/s obtained for $t_{MOT}$ = 4~ms ($t_{\ell}$ = 7 ms).

\begin{figure}

\includegraphics[width=\columnwidth]{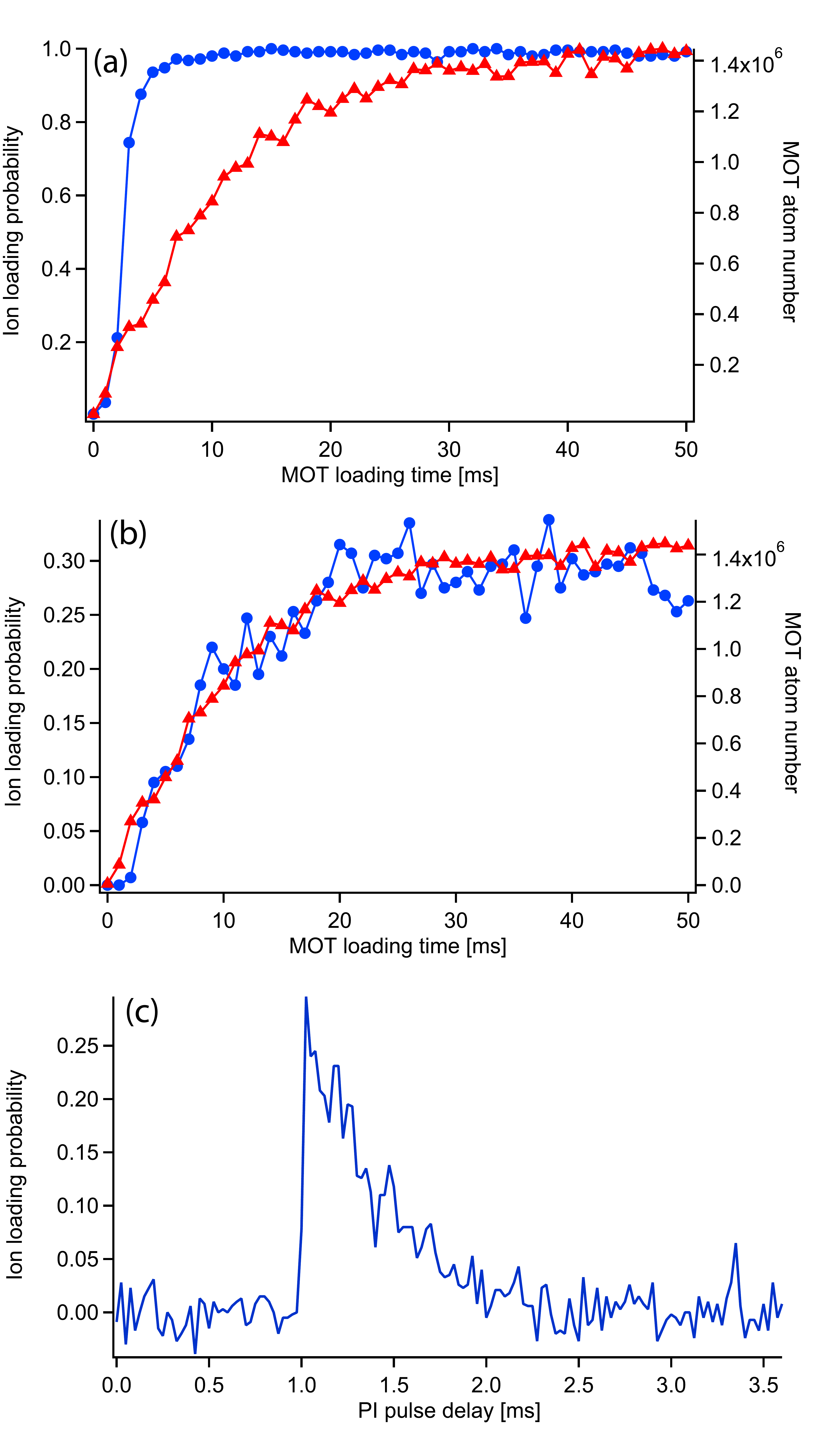}

\caption{\label{fig:satvsunsat} (Color online) Ion loading in saturated and unsaturated regimes.  (a) Ion loading probability (circles, left axis) and MOT atom number (triangles, right axis) as a function of MOT loading time showing the ion loading probability (ILP) saturating before the MOT atom number.  Here the photoionization (PI) lasers are on for 3 ms after the MOT beams are turned off.  (b) Same as (a), except the first-step PI laser is pulsed on for only 10 $\mu$s for the ion loading data (circles, left axis).  In this case, the ILP is not saturated and is proportional to the MOT atom number (triangles, right axis).  Lines in (a) and (b) are guides to the eye. (c) Unsaturated ILP as a function of delay of the first-step PI laser pulse after the MOT beams are shut off, with a fixed pulse width of 25 $\mu$s and a fixed MOT loading time of 10 ms.  The ILP is here proportional to the temporal density profile of the pulsed cold atomic beam.  A background loading rate due to first-step PI excitation from the push beam (which is on continuously for 5 ms after the MOT beams are shut off) has been subtracted. All data are taken with  $T_\mathrm{oven}=425^{\circ}$C and a MOT beam intensity of 40 mW/cm$^2$.}

\end{figure}

While it is expected that the ion loading probability should saturate with $t_{MOT}$ due to the MOT atom number saturation, we observe that $P_{ion}$ saturates at a time significantly shorter than $\tau_{MOT}$, as shown in Fig.~\ref{fig:satvsunsat}(a).  This is a direct result of our inability to load and detect more than one ion in the trap; once the cold atom flux is high enough so that the probability to load one ion becomes appreciable, the probability to load another becomes small.

To verify this, we modify our loading sequence to purposely reduce the loading probability.  We do this by pulsing on the 461 nm PI beam for only a short duration, delayed by 1.2~ms after the push beams turn on, after explicitly verifying that the loading probability is linear in a range around our chosen pulse duration.  We then measure the ion loading probability versus $t_{MOT}$ in this unsaturated regime. The data are shown in Fig.~\ref{fig:satvsunsat}(b) for an oven temperature of 425$^{\circ}$C, a MOT beam intensity of 40 mW/cm$^2$, and a PI pulse duration of 10 $\mu$s.  We see that the ion loading curve's shape matches well with that of the MOT loading curve, indicating that the ion loading rate is indeed proportional to the cold atom number in this regime.  From this we can infer that we could load ions at a higher rate than the one we observe when using the long 3 ms PI pulse if our trap were capable of stably trapping multiple ions.

We can quantify this by measuring the ion loading rate as a function of the 461 nm PI pulse delay from the time the MOT beams turn off (and the push beam turns on), using a short PI pulse so that we are in the unsaturated ion loading regime.  The data from this measurement are shown in Fig.~\ref{fig:satvsunsat}(c). We use a PI pulse duration of 25~$\mu$s and vary the pulse delay in steps of 25~$\mu$s working at a fixed  value of $t_{MOT}$ = 10 ms.  The oven temperature and MOT beam intensity are 425$^{\circ}$C and 40~mW/cm$^2$, respectively.  Since the ion loading probability is proportional to the cold atom number, this measurement gives the temporal density profile of the incoming atom pulse.  We see that the atoms arrive in a time as early as 1.0~ms, with the center of mass of the pulse arriving in 1.4 ms.  Furthermore, because the value of each data point in the temporal profile represents the probability to load one ion in a 25~$\mu$s time slice, we can sum the data and interpret the result as the expectation value of the number of ions loaded in the entire 3.5~ms window.  Doing this, we obtain a value of 4.7 from which we infer that this average number of ions could be loaded in $t_{\ell}=13.5$~ms into a trap capable of stably trapping many ions.  This implies an effective unsaturated ion loading rate of 350~ions/s.

The discussion of this effective rate is meant to quantitatively illustrate the role saturation plays in our loading process and is not an actual rate that we attain in these experiments.  However, it shows that higher rates should be achievable if saturation can be mitigated or avoided.  Such would be the case for loading an array of traps where the density of the traps in the array is comparable to the cold atomic beam density.  In this configuration, we estimate that we could load a $\sim10$~$\mu$m~pitch, 2D array of order $10^5$ sites from one cold atomic beam pulse in less than 10~ms.

The measured loading rate of 125 ions/s is remarkable in that it is among the highest reported in the literature in spite of loading into an extremely shallow trap that can stably trap only a single ion.  However, it is difficult to compare the rate directly to that which has been achieved in other work because the rate depends on experimental parameters such as the atomic source flux, the PI laser intensities, and the particular trap used.  We note, however, that while loading at a high rate itself is of great importance, a primary goal of our work is to demonstrate such a rate with a technique that minimally deposits material on the trap electrodes.

To provide insight into the merit of our technique, it is useful to first compare it to the method of loading our trap using direct PI of atoms in a hot atomic beam, which is the technique most commonly used.  While we do not experimentally demonstrate loading from a hot beam in this work, a comparison can still be made.  If we were to load directly from the hot beam generated from a strontium oven at 425$^{\circ}$C into our $E_{trap}$ = 15 meV deep trap, we find, from integration of the energy distribution function for atoms in a beam \cite{Ramsey:MolBeams} up to $E_{trap}$, that only $\sim$ 3$\%$ of the atoms can be trapped.  Comparing this to 100$\%$ of our 2.3 meV cold accelerated atoms that can be trapped, we see that at a minimum we achieve more than a factor of 30 improvement in the amount of unwanted deposition on the trap chip for a given amount of atom flux.  If $E_{trap}$ were as shallow as our 2.3 meV accelerated atom energy, then the improvement in unwanted deposition would be greater than three orders of magnitude.  Furthermore, we note that our cold atom source is isotopically pure so that if loading of a single isotope with abundance  $a_i$ were desired, we would observe an additional improvement by a factor of  $a_i^{-1}$.  For example, for loading $^{87}$Sr ($a_{i}=0.07$) or $^{43}$Ca ($a_{i}=0.0014$), another factor of 14 or 700 improvement is expected.

We note that if one desired to load into deep traps ($E_{trap}/k_B>T_{oven}$), PI directly from a hot atomic beam may ultimately lead to higher loading rates since the flux can be easily increased by many orders of magnitude with only a slight increase in oven temperature \cite{Oxford:IsotopeSelectLoading:PRA:04}.  Additionally, if deposition on the trap chip is of little concern, loading rates even into shallow traps may again be much greater for loading directly from the hot atomic beam as compared with our technique, again because the hot flux can easily be increased.  However, at some point we suspect that the loading rate will saturate as the hot atomic beam flux is increased due to collisions between the ions and the hot atoms, though at what flux this occurs is not known.  Regardless, for loading shallow, isotopically pure traps cleanly, our technique should always prove superior.

It is difficult to directly compare the ion loading rate using our method to the rate obtained using the technique of loading from a collocated MOT \cite{Chuang:MOTionload:PRA:07}.  In the latter technique, much higher rates were reported but ions were loaded into a much larger and deeper trap and the reported rate did not take into account the time it took to load the MOT.  However, we note that our method compares favorably with regard to surface contamination as loading from the collocated MOT necessitates exposure of the ion trap to the hot atom source and therefore does not mitigate the problem of substantial deposition of material onto the trap electrodes.

The technique of backside loading, where neutral atoms from a hot oven are delivered through a 100~$\mu$m-scale hole from the back of the ion trap chip and are subsequently ionized, prevents deposition onto the trap electrodes; however, the machining of these holes greatly complicates the trap fabrication and limits the choice of substrates that can be used, which may be important for reduction of possibly material-dependent anomalous heating of ions.  Furthermore, backside loading is not likely to be a scalable solution capable of loading large 2D arrays of small, densely packed traps.   To do this would require the highly non-trivial engineering of an array of small loading holes at a density comparable to the trap array itself or else using a small number of remote loading zones.  The latter technique would suffer from a significantly slower loading rate, as delays would be incurred due to the need to shuttle ions to individual sites from remote loading locations \cite{NIST:IonRacetrack:NJP:10}.

\section{Isotopic Selectivity}
While we do not perform a detailed study of the isotopic selectivity of our loading process in these experiments, we argue that loading from a remote, pre-cooled source of neutral atoms offers the potential for improved selectivity over other demonstrated techniques.  Indeed, over the course of the experiments presented here, we have witnessed the loading of thousands of $^{88}$Sr ions, imaged with the CCD camera, without observing evidence of loading of a single ion of a different isotope \cite{Isotopenote}.  The MOT provides a high level of isotopic filtering in our case and further isotopic filtering is expected from the resonant push beam, which selectively accelerates the desired isotope toward the ion trap.  We note that other stages of isotopic filtering can be performed by pushing any undesired isotopes out of the MOT and away from the ion trap with additional resonant lasers or through time-of-flight methods whereby the photoionization lasers are switched on only when the desired isotope is present in the ion trap region. Although we did not implement these other filter stages in this work, it is a straightforward improvement that can be easily added to the ion loading sequence with minimal effect on the loading rate.

Photoionization of the cold atoms with lasers traveling orthogonally to the atomic beam provides another stage of isotopic selectivity.  Although it is common to techniques implementing loading from a hot atomic beam, it should be more selective in our approach because of a reduction in the Doppler-broadened linewidth resulting from the cold atoms' relatively lower transverse velocity.  The additional isotopic selectivity is welcome, but we note that a major advantage of our technique is that one may not need to rely on PI at all to perform the selection if the cold beam is already sufficiently isotopically pure.  This enables an increase in PI rate (and corresponding ion loading rate) through an increase in the PI laser intensities without sacrificing isotopic selectivity due to power broadening.  Additionally, for loading large arrays of ions, one may desire to photoionize atoms from a large spatial region of the atomic beam; however, one loses isotopic selectivity when ionizing atoms positioned away from the central atomic beam axis because these atoms are moving with higher transverse velocity and Doppler broadening will be an issue \cite{Oxford:IsotopeSelectLoading:PRA:04}.   Our technique of PI loading from an already isotopically pure beam circumvents this issue and therefore allows for substantially higher loading rates into large arrays without decreased isotopic selectivity.

The remote nature of our loading source adds an additional level of isotopic purity as it offers protection against charge exchange between a trapped ion of a desired isotope with an atom of an undesired isotope \cite{Oxford:IsotopeSelectLoading:PRA:04, ChuangVuletic:YbChrgEx:PRL:09}.  Since the cold beam is isotopically pure and no atoms from the hot oven can reach the ion trap region, there is no chance for charge exchange to occur between different isotopes.   This is in contrast even to the work of Ref. \onlinecite{Chuang:MOTionload:PRA:07} where an isotopically pure MOT was used for ion loading.  In that work the trap region was exposed to an isotopically impure hot atomic beam due to the collocation of the MOT and ion trap and, as a result, charge exchange could still occur.

\section{Conclusion}
We have demonstrated a technique utilizing a remote, pre-cooled neutral source of atoms for loading ions that is ideally suited for loading at high rates into low-depth traps with high isotopic purity.  This loading is achieved with minimal deposition of metal on the trap electrodes due to the remote nature of the cold atom source.  Our ion trap is maintained at a cryogenic temperature during loading, demonstrating that the technique is compatible with low temperature systems.  We have demonstrated a loading rate as high as 125 ions/s into a 15 meV deep trap.  We note that although we chose $^{88}$Sr for these experiments, the technique is general with respect to ion species or isotope; it can be applied to any ion whose neutral version can be laser cooled and trapped (e.g., various isotopes of Mg, Ca, Sr, Ba, and Yb) \cite{87SrMOTnote}.  With multiple stages of isotopic filtering, we expect that our method can lead to loading of large arrays of ions with unsurpassed isotopic purity.  While the degree of isotopic purity will be ion species-dependent due to differences in isotope shifts and relative natural abundances, our method should always offer superior selectivity to other reported loading methods for a particular ion of choice.

We note that the only major addition to the apparatus required to implement our loading method as compared with direct loading from a hot atomic beam is a high power ($\sim 100$ mW) laser for the MOT; however, this same laser will replace the lower power version used by most researchers to do the first step of PI.  As a result, the loading method discussed here offers substantial improvement over other methods while requiring little additional complication.

The minimal surface contamination resulting from this method, together with the cryogenic operation of our apparatus, makes our system well-suited to the task of reducing anomalous heating rates in ion traps  \cite{Chaung:CryoHeatRates:PRL08}.  With such improvements, traps may be made smaller without incurring the cost of lower quantum gate fidelity.  Two-dimensional arrays of traps can then be made to take full advantage of the high loading rate attainable due to the high phase-space density of our cold atom source.

\begin{acknowledgments}
We thank Daniel Baker, Valdimir Bolkhovsky, Peter Murphy, and Jacob Zwart for technical assistance with ion trap fabrication and packaging. This work is sponsored by the Assistant Secretary of Defense under Air Force Contract $\#$FA8721-05-C-0002. Opinions, interpretations, conclusions and recommendations are those of the authors and are not necessarily endorsed by the United States Government.
\end{acknowledgments}

\bibliography{IonLoading_PRAfinal}

\begin{thebibliography}{42}
\bibitem [{MOT()}]{MOTDensitynote}%
  \BibitemOpen
  \href@noop {} {}\bibinfo {note} {We note that for our case where the ion trap
  region is smaller than the atom beam diameter at the trap location, the
  density of the beam is the important figure of merit since this determines
  the number of atoms that spatially overlap with the trap region; however, the
  MOT density increases nearly linearly with the atom number in MOTs of the
  size obtained in this work \cite{Ketterle:darkSPOT:PRL:93}. Additionally, the
  final beam diameter is determined almost completely by the transverse
  spreading of the atoms as they are pushed, which is a function of the
  constant push beam intensity and MOT temperature. As a result, the density
  should increase nearly linearly with MOT atom number.}\BibitemShut {Stop}%
\bibitem [{Iso()}]{Isotopenote}%
  \BibitemOpen
  \href@noop {} {}\bibinfo {note} {Evidence of loading of an undesired isotope
  would consist of a notably higher measured fluorescence rate in the case of
  $^{86}$Sr (for the 422 nm Doppler laser detuning used) or the presence of a
  dark ion when multiple ions are simultaneously loaded in the case of
  $^{87}$Sr.}\BibitemShut {Stop}%
\bibitem [{87S()}]{87SrMOTnote}%
  \BibitemOpen
  \href@noop {} {}\bibinfo {note} {We observe loading of a $^{86}$Sr MOT for a
  different choice of MOT laser frequency in these experiments. Although ion
  loading was not investigated here using this MOT as the source, the ability
  to trap a different neutral isotope in the same apparatus demonstrates the
  potential generality of our ion loading method.}\BibitemShut {Stop}%
\end{thebibliography}

\end{document}